\documentclass[conference]{IEEEtran}
\IEEEoverridecommandlockouts
\usepackage{cite}
\usepackage{amsmath,amssymb,amsfonts}
\usepackage{algorithmic}
\usepackage{graphicx}
\usepackage{textcomp}
\usepackage{xcolor}

\usepackage{verbatim}
\usepackage{enumerate}
\usepackage{float}
\usepackage{hyperref}
\usepackage{float}
\usepackage{caption}
\usepackage{subcaption}

\def\BibTeX{{\rm B\kern-.05em{\sc i\kern-.025em b}\kern-.08em
    T\kern-.1667em\lower.7ex\hbox{E}\kern-.125emX}}
\begin{document}

\title{Reducing the Complexity of the Sensor-Target Coverage Problem Through Point and Set Classification
}

\author{\IEEEauthorblockN{1\textsuperscript{st} Christopher Thron}
\IEEEauthorblockA{\textit{Department of Science and Mathematics)} \\
\textit{Texas A\& M University-Central Texas)}\\
Killeen, TX USA \\
thron@tamuct.edu}
\and
\IEEEauthorblockN{2\textsuperscript{nd} Anthony Moreno}
\IEEEauthorblockA{\textit{Department of Science and Mathematics)} \\
\textit{Texas A\& M University-Central Texas)}\\
Killeen, TX USA \\
thron@tamuct.edu}
}

\maketitle

\begin{abstract}
The problem of covering random points in a plane with sets of a given shape has several practical applications in communications and operations research. One especially prominent application is the coverage of randomly-located points of interest by randomly-located sensors in a wireless sensor network.
In this article we consider the situation of a large area containing randomly placed points (representing points of interest), as well a number of randomly-placed disks of equal radius in the same region (representing individual sensors' coverage areas). The problem of finding the smallest possible set of disks that cover the given points is known to be NP-complete. We show that the computational complexity may be reduced by classifying the disks into several definite classes that can be characterized as necessary, excludable, or indeterminate.  The problem may then be reduced to considering only the indeterminate sets and the points that they cover. In addition, indeterminate sets and the points that they cover may be divided into disjoint ``islands'' that can be solved separately. Hence the actual complexity is determined by the number of points and sets in the largest island. We run a number of simulations to show how the proportion of sets and points of various types depend on two basic scale-invariant parameters related to point and set density. We show that enormous reductions in complexity can be achieved even in situations where point and set density is relatively high. 
\end{abstract}


\section{Introduction}\label{sec.Intro}

\subsection{Overview}
The problem of covering a given set of points with a given collection of sets can be solved exactly using integer  programming \cite{r1}.   However, the computation is known to be NP-complete \cite{r2}. Several approximate algorithms have been developed, some of which use Monte Carlo methods \cite{r3}. Before applying these methods, it is advantageous to reduce the size of the problem as much as possible. Some reductions are quite simple to compute: for example some sets contain no points, and can be removed. Some points are covered by a single set, and those single-covering sets must necessarily be in any cover. Other sets may be included in the union of necessary sets, and these also may be excluded from the minimal cover.  Finally, it may be possible to break the simplified cover problem into disjoint problems that can be solved separately.
In this paper we explore how much reduction is possible, and how this depends on the initial parameters used to generate the random points and disks.
We specialize to the case of circular cover sets, but we expect that results may apply  qualitatively to other types of sets.  
Our  results also apply in cases where different sets have different costs, since the classification of sets as necessary or unnecessary does not depend on the cost function. 

This problem can be recast as a problem involving a certain type of random bipartite graph. Disks and points correspond to nodes in two different classes, while edges are drawn between disk nodes and the points they contain. 
Random graph theory was originated by Paul Erd{\H{o}}s and Alfr{\'e}d, R{\'e}nyi \cite{r7}, and since then has developed into an extensive theory \cite{r8}. A main result of Erd{\H{o}}s and R{\'e}nyi
is the existence of phase transitions, which are sudden changes in the frequencies of certain types of subgraphs depending on the proportion of edges that are included in the graph \cite{r8}.  

\subsection{Definition of parameters}

Given a dataset of uniformly-distributed random points in a square region of the plane and a set of equally-sized disks in the plane with uniformly-generated centers, we define the following parameters:
\begin{align*}
N &= \text{number of sets (disks) in dataset}  \\
M &= \text{number of points in dataset}  \\
A &= \text{total area of the square region}  \\
a &= \text{area of each set in dataset}.
\end{align*}
From these basic parameters we may derive two dimensionless parameters:
\begin{align*}
\gamma &\equiv \frac{Ma}{A} \text{\quad (max point area fraction)}\\  
\phi &\equiv \frac{Na}{A} \text{\quad (max set coverage fraction)} 
\end{align*}
These parameters reflect the point and set density respectively. 
A value of $\gamma=1$ means that the average number of points within any area of size $a$ is 1. Similarly a value of $\phi=1$ means that a random point within the region is covered by one set on average. These parameters are scale-invariant: for any large rectangle containing random points and sets with a particular value of $\gamma$ and $\phi$, any  rectangle within that rectangle will have the same parameters.

For very small values of $a/A$ the boundary effects can be neglected. In this case, the parameters $\gamma$ and $\phi$ 
are sufficient to completely characterize the behavior of the system, in the sense that proportions of points and sets
with certain incidence properties depend on the basic parameters only through $\phi$ and $\gamma$.

\subsection{Characterization of points and sets}
We may identify 4 kinds of points:
\begin{itemize}
\item	Uncovered points: points that are not in any of the given sets in the cover;
\item	Single-covered points: points that are in exactly one set;
\item	Collateral points: points that are in the same set as a single-covered point;
\item Indeterminate points: all other points. 
\end{itemize}

We may also identify  five  kinds of sets:
\begin{itemize}
\item	Non-covering sets: sets that do not cover any points;
\item	Single-covering sets: sets that contain a single-covered point, and thus must be included in any cover;
\item Collateral sets: sets that  contained only points that are already contained within the union of single-covering sets, and thus do not cover any additional points;
\item	Indeterminate sets: all other sets.
\end{itemize}

We shall see that the point and set classifications are complicated by the fact that they are determined by an iterative process. After the first round of single-covered and  collateral points have been identified, additional redundant sets may be found and removed, which gives rise to additional single-covered and collateral points.

When finding an optimal set cover of the given points, single-covered points and single-covering sets must necessarily be in the cover. Non-covered points will not be in any subcover, and thus may be removed. Similarly, non-covering sets are useless and can be removed. Collateral points are automatically covered by single-covering sets. The problem can thus be reduced to  finding an optimal cover of indeterminate points with indeterminate sets. 
Hence the number of indeterminate points and sets provide an upper bound to the practical complexity of the coverage problem.  

This upper bound may be further reduced by partitioning indeterminate sets and points  into  \emph{islands}, where an island is a minimal collection of indeterminate sets (together with the points that they cover)  such that they do not intersect with any indeterminate set outside of the island.
Each island can be solved independently to get the overall optimal solution. Thus the maximum island size (including both the number of points and the number of sets) is the true measure of the practical computational complexity of the problem.

It follows that characterizing the proportions of different types of points and sets provides us with key insight into the essential behavior of this system. In this ARTICLE, we give heatmap representations of proportions obtained by simulation.

\section{Methods} \label{Experiments}

In this section, we describe simulations we performed to estimate point and set proportions as a function of the point and set coverage fraction parameters $\gamma$ and $\phi$.
All code for these simulations was written in the \texttt{Python} programming language (version 3.7.3 64-bit) and executed in the Scientific Python Development Environment  (a.k.a \texttt{Spyder}) version 4.1.5.  Numerous functions from both the \texttt{NumPy} (version 1.16.4) and \texttt{Matplotlib} libraries (version 3.1.0) are used. A more complete code description may be found in \cite{r9}, and the source code is available from the corresponding author on request.
 
In the simulations, the ranges of $\gamma$ and $\phi$ were both set as $3 \le \gamma, \phi \le 12$. These ranges were determined by preliminary investigation, which indicated that this range showed the most interesting behavior.  

A rescaling of $M$, $N$ and $a$ was used to reduce simulation time. 
The rescaling preserved the values of $\gamma$ and $\phi$, so the calculated proportions shown in the heatmaps were not affected.
The rescaling was arranged so that $\min(M,N) = 1000$.  

For each value of $M$ and $N$, 105 configurations were run, and for each configuration the 
following quantities were recorded:

\begin{enumerate}
\item $M$, $N$ and $a$;
\item Number of uncovered, single-covered, collateral, and indeterminate points;
\item  Number of non-covering, single-covering, collateral, and indeterminate sets;
\item Number of indeterminate points and sets in the largest island
\item Standard deviation of number of indeterminate points and sets per island
\end{enumerate}

All proportions were averaged over 105 random configurations with the same values of  $\gamma$ and $\phi$. Similarly, all standard deviations were computed based on the same 105 configurations for each set of values of  ($\gamma$, $\phi$).

\section{Results} \label{Results}
\begin{figure}[H]
	\centering
	\includegraphics[width=3in]{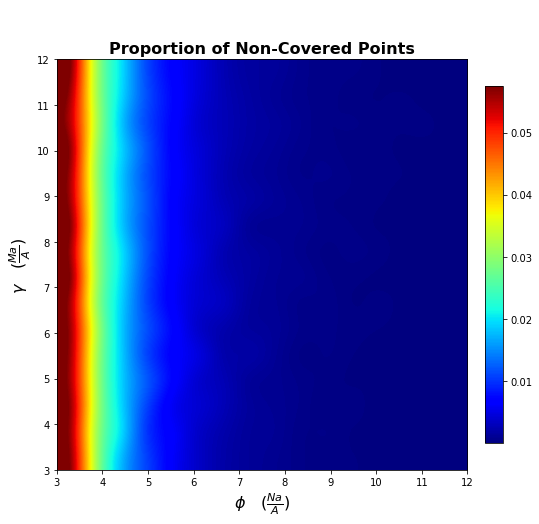}
	\caption{Experimental proportion of uncovered points.}\label{fig:Exp-ncvPnts}
	\end{figure}

\begin{figure}[H]
\centering
	\centering
	\includegraphics[width=3in]{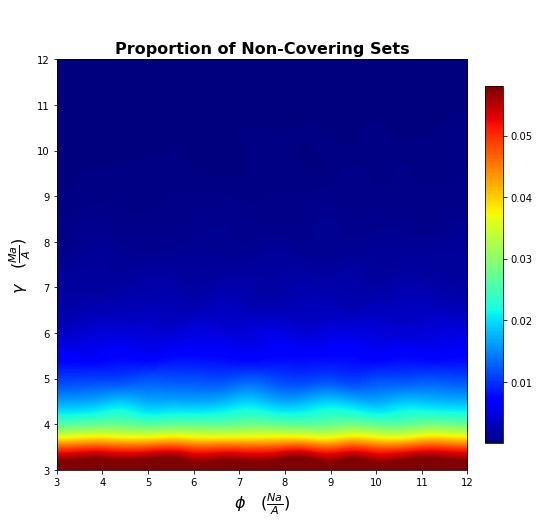}
	\caption{Experimental proportion of non-covering sets}
	\label{fig2}
\end{figure}

Figures~\ref{fig:Exp-ncvPnts} and \ref{fig2} give simulation results for the proportions of uncovered points and non-covering sets as a function of $\gamma$ and $\phi$.  The graphs show that more than  99\% of the points are covered by randomly-placed disks as long as  $ \phi > 5.5$; and similarly more than 99\% of disks cover at least one point as long as $\gamma > 5.5$. The graphs demonstrate the symmetry between  These results agree closely with theoretical expressions derived in \cite{r9}.

\begin{figure}[H]
\centering	
\includegraphics[width=3in]{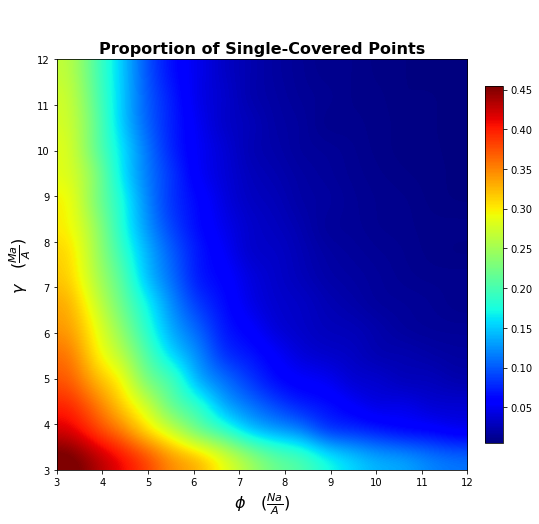}
\caption{Experimental proportion of single-covered points}
	\label{fig3}
\end{figure}

\begin{figure}[H]
\centering
	\includegraphics[width=3in]{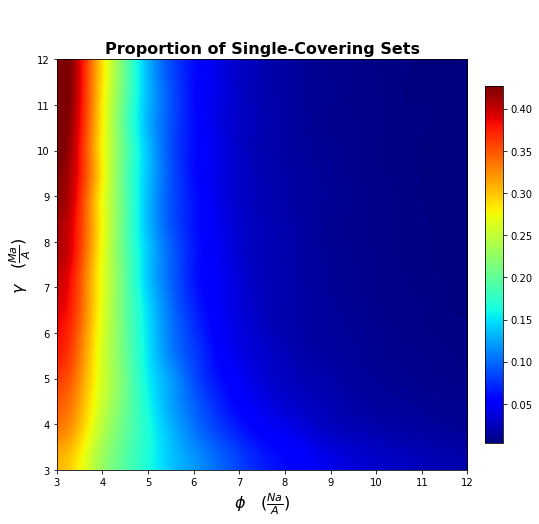}
\caption{Experimental proportion of single-covering sets
	\label{fig4}}
\end{figure}

Figures~\ref{fig3} and \ref{fig4} give simulated values for the proportions of single-covered points and sets respectively as functions of $\gamma$ and $\phi$. Single-covering sets are necessary to the cover, and thus may be removed when trying to find an optimal cover. This can lead to considerable reductions in complexity: for example, when $\phi = 5.5$ about 20\% of sets are single-covering  and between 10-40\% of points are single-covered, regardless of the point density.

\begin{figure}[H]
\begin{center}
\includegraphics[width=3in]{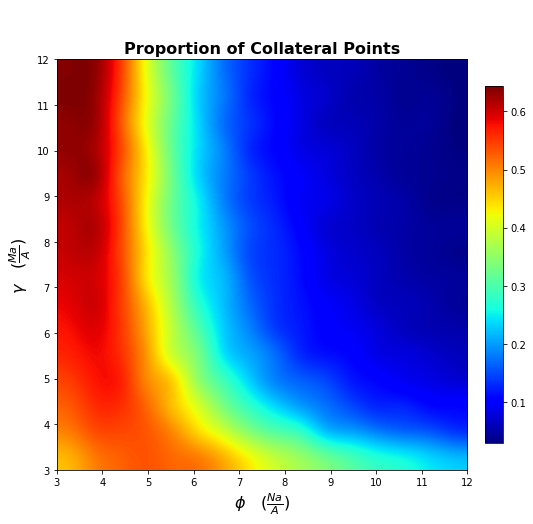}
\caption{Proportion of collateral points - experimental results}
\label{fig:colPnts}
\end{center}
\end{figure}

\begin{figure}[H]
\begin{center}
\includegraphics[width=3in]{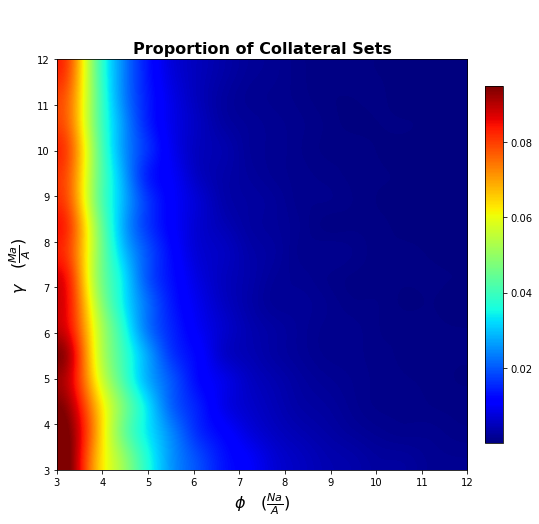}
\caption{Proportion of collateral sets - experimental results}
\label{fig:colSets}
\end{center}
\end{figure}

 Figures~\ref{fig:colPnts} and \ref{fig:colSets} 
show the experimentally measured proportions of collateral points and collateral sets, respectively. As explained in Section~\ref{sec.Intro},  collateral sets are excluded from the optimal cover, and collateral points may be excluded from the calculation of the optimal cover because they are included in single-covering sets. The proportion of collateral points is surprisingly high,about 40\% when $\phi=5.5$. Once again this  leads to reductions in complexity when computing the optimal cover.

\begin{figure}[H]
\begin{center}
\includegraphics[width=3in]{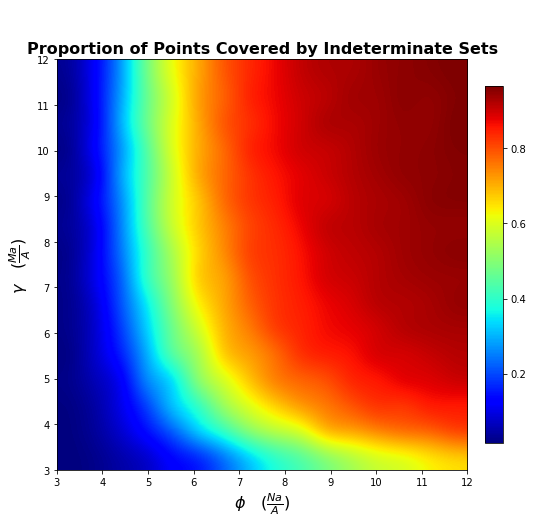}
\caption{Experimental proportion of indeterminate points.\label{fig:indPnts}}
\end{center}
\end{figure}

Figures~\ref{fig:indPnts} and 
show the experimentally measured proportions of indeterminate points and sets respectively, which are the remaining points and sets that serve as inputs to the point-set covering algorithm. As expected, the proportions becomes larger with increasing $\gamma$ and $\phi$. However, even for rather large values of point and set density, the proportion can still be below 0.5. We see for example if $\gamma=\phi=6$, the proportion of indeterminate points is roughly 0.5.  

\begin{figure}[H]
\begin{center}
\includegraphics[width=3in]{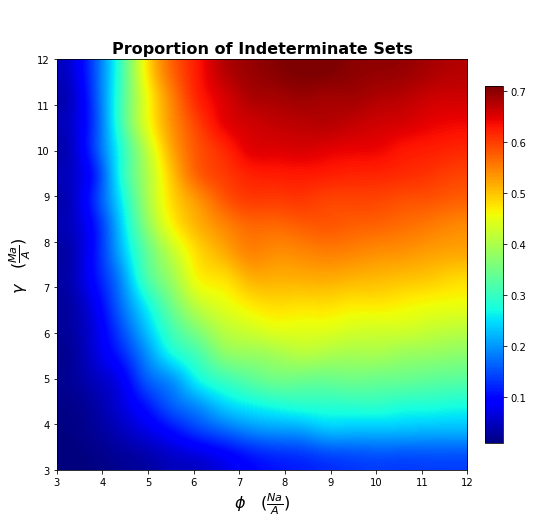}
\caption{Experimental proportion of  indeterminate sets.\label{fig:indSets}}
\end{center}
\end{figure}

\begin{figure}[H]
\begin{center}
\includegraphics[width=3in]{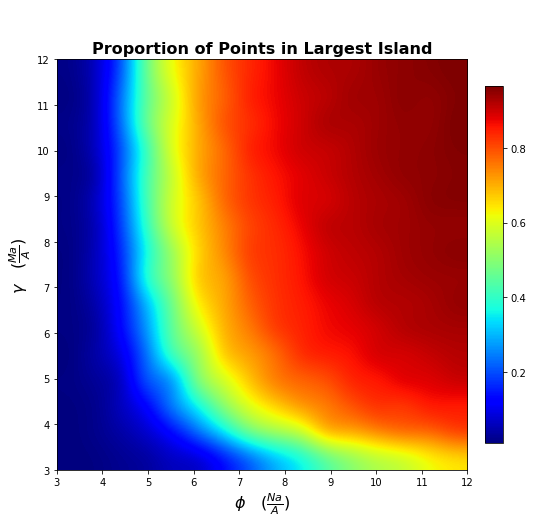}
\caption{Experimental proportion of points in largest island}
\label{fig:Isl_Pnts}
\end{center}
\end{figure}

\begin{figure}[H]
\begin{center}
\includegraphics[width=3in]{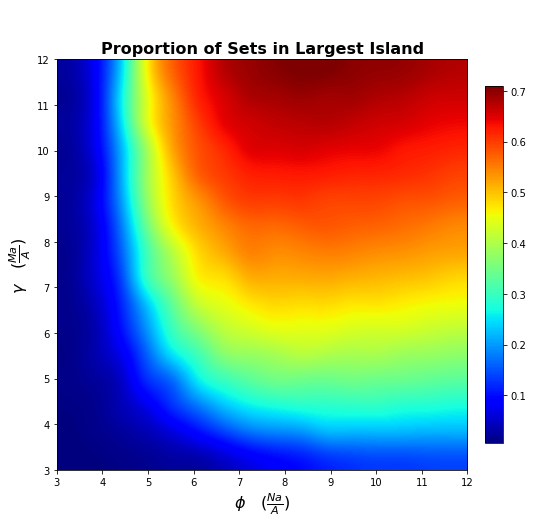}
\caption{Experimental proportion of sets in largest island}
\label{fig:Isl_Sets}
\end{center}
\end{figure}

Figures~\ref{fig:Isl_Pnts} and \ref{fig:Isl_Sets}  show the proportion of all points that are in the largest island. This figure closely resembles Figure~\ref{fig:indPnts}, which indicates that most indeterminate points and sets belong a single island. As explained in Section~\ref{sec.Intro}, the  numbers of points and sets in the largest island determine the complexity of the algorithm, because solving the coverage problem for this island is the largest subproblem which may be solved independently of the subproblem posed by other islands.  

\section{Discussion and future investigations}
Our simulations show that even for fairly high point and set densities,  the complexity of the set cover problem be enormously reduced through point and set classification. Note that since complexity increases exponentially, a reduction in both points and sets by 50\% (which is obtained when $\phi \approx 6$ and $\gamma \approx 7$ represents a reduction of 50\% in log complexity, which means the complexity is reduced to the square root of its former value.  Furthermore, there are very steep gradients in Figures~\ref{fig:Isl_Pnts} and \ref{fig:Isl_Sets}  when $\phi \approx 6$. This means that reducing the number of indeterminate sets only slightly by selecting a few high-incidence sets as part of the cover, the complexity will be further reduced and the remaining problem may possibly be solved exactly.
 Future work may focus on working out practical details for such a complexity-reducing algorithm. In addition, future investigations may be conducted into finding analytical expressions (either exact or approximate) for the contours in the figures in Section~\ref{Results}.


\begin{thebibliography}{00}

\bibitem{r1}, V. Vazirani, ``Approximation algorithms,'' Springer, 2001.

\bibitem{r2} T. H. Cormen, C.E. Leiserson, R. L. Rivest. ``Introduction to
Algorithms''. The MIT Press, 1991.
\bibitem{r3} Q. Yang, A. Nofsinger, J. McPeek, J. Phinney, \&  R. Knuesel, ``A complete solution to the set covering problem,'' In Proceedings of the International Conference on Scientific Computing (CSC) (p. 36). The Steering Committee of The World Congress in Computer Science, Computer Engineering and Applied Computing (WorldComp), 2015.
\bibitem{r4} J. E. Beasley,  and K.  Jörnsten. ``Enhancing an algorithm for set covering problems,'' European Journal of Operational Research, 58(2), 293-300, 1990. 
\bibitem{r5} S. Haddadi. ``Simple Lagrangian heuristic for the set covering problem,'' European Journal of Operational Research, 97(1), 200-204, 1996.
\bibitem{r6} L. W. 
Jacobs and  M. J. Brusco.  ``Note: A local‐search heuristic for large set‐covering problems,'' Naval Research Logistics (NRL), 42(7), 1129-1140, 1995.
\bibitem{r7}
P. Erd{\H{o}}s and A. R{\'e}nyi,
``On random graphs I,''
Publicationes Mathematicae (Debrecen) 6, 290--297,
1959.\bibitem{r8}
A, R{\'e}ka and A.-L. Barab{\'a}si, ``Statistical mechanics of complex networks,'' Reviews of Modern Physics 74(1),
 47, American Physical Society, 2002.
 \bibitem{r9}
 C. Thron and  A. Moreno, ``Covering Random Points in the Plane with Equal-Sized Disks: Theory and Simulations. Available at SSRN, \url{https://ssrn.com/abstract=375825} or \url{http://dx.doi.org/10.2139/ssrn.3758254}, 2020.
 

\end{thebibliography}
\end{document}